\documentclass{article}[12pt]
\usepackage{amsmath,amsthm,amsfonts,amssymb,amscd, latexsym, fullpage}

\begin{document}
\theoremstyle{plain}
\newtheorem{thm}{\sc Theorem}
\newtheorem{lem}[thm]{\sc Lemma}
\newtheorem{cor}{\sc Corollary}[section]
\newtheorem{prop}[thm]{\sc Proposition}

\theoremstyle{remark}
\newtheorem{Case}{\bf Case}[section]
\newtheorem{rem}[thm]{\sc Remark}

\theoremstyle{definition}
\newtheorem{Def}{\bf Definition}[section]
\newtheorem*{pf}{\sc Proof}

\theoremstyle{definition}
\newtheorem{Conj}{\bf Conjecture}[section]

\newcommand{\N}{{\mathbb N}}
\newcommand{\ve}{{\varepsilon}}
\newcommand{\be}{{\beta}}
\newcommand{\la}{{\lambda}}
\newcommand{\teta}{{\theta}}
\newcommand{\La}{{\Lambda}}
\newcommand{\ga}{{\gamma}}
\newcommand{\Ga}{{\Gamma}}
\newcommand{\po}{{\partial}}
\newcommand{\ov}{\overline}
\newcommand{\re}{{\mathbb{R}}}
\newcommand{\Pc}{{\mathbb{P}}} 
\newcommand{\dc}{{\mathcal D}}
\newcommand{\na}{{\mathbb N}}
\newcommand{\co}{{\mathbb C}}
\newcommand{\fc}{{\mathcal F}}
\newcommand{\al}{{\alpha}}
\newcommand{\lc}{{\mathcal L}}
\newcommand{\om}{{\omega}}  
\newcommand{\Z}{{\mathbb Z}}
\newcommand{\rth}{{\mathbb{R}^3}}
\newcommand{\W}{{\mathcal W}}  
\renewcommand{\H}{{\mathcal H}}
\newcommand{\cc}{{\mathcal C}}
\newcommand{\E}{{\mathcal E}}
\newcommand{\PP}{{\mathcal{P}}}
\newcommand{\M}{{\mathcal M}}
\newcommand{\Si}{\Sigma}

\newcommand{\inv}{{^{-1}}}

\begin{title}
{Explicit soliton-black hole correspondence for static configurations}
\end{title}

\begin{author}
{Shabnam Beheshti\footnote{beheshti@math.umass.edu} and Floyd L. Williams\footnote{williams@math.umass.edu}\\
Department of Mathematics and Statistics\\
University of Massachusetts\\
Amherst, MA 01003}
\end{author} 

\maketitle

\begin{abstract}
We construct an explicit map that transforms static, generalized sine-Gordon metrics to black hole type metrics.  This, in particular, provides for a further description of 
the Cadoni correspondence (which extends the Gegenberg-Kunstatter
correspondence) of soliton solutions and extremal black hole solutions
in 2D dilaton gravity.\\
\vspace{.5 cm}
\end{abstract}

\section{Introduction}
An interesting, intriguing connection between Euclidean N-soliton
sine-Gordon solutions and Lorentzian black hole solutions in
Jackiw-Teitelboim dilaton gravity has been 
established by J. Gegenberg and G. Kunstatter ~\cite{4,5,6}.  In case
N=1, a concrete transformation was constructed that explicates this
connection ~\cite{9}; see also ~\cite{10,11}.  The 
construction of such a transformation in general seems to be a difficult problem as it involves, in particular, finding explicit solutions of a system of dilaton field 
equations.  Recently, these field equations were solved for a kink-antikink soliton (similar solutions were found in ~\cite{5}),  and thus an explicit transformation was also 
constructed in this case ~\cite{2} that further implements the work in
~\cite{4,5}.

We present a transformation $\Psi$ that takes \underline{any} generalized \underline{static} sine-Gordon type metric to a black hole type metric.  In particular we present 
additional solutions of the dilaton field equations (even in the non-static case) and a further description of the M. Cadoni correspondence ~\cite{3} between extremal black holes 
and generalized solitons--generalized sine-Gordon solutions that \underline{do} \underline{not} give rise to constant curvature space-times, as in the Gegenberg-Kunstatter discussion.

We dedicate this paper to the memory of Professor Melvyn
Berger--friend, and outstanding scholar in non-linear phenomenon.

\section{Field equations for 2D dilaton gravity}
Given a potential function $V(r)$ and $l>0$, we consider the general two-dimensional dilaton gravity theory with action integral
\begin{equation}
 I(\tau,g)=\frac{1}{2G}\int d^2x \sqrt{-g}\left( \tau R(g) + \frac{V
  \circ \tau}{l^2} \right),
\end{equation}
for which the equations of motion are
\begin{eqnarray}
R(g) + \frac{V' \circ \tau}{l^2} &=& 0 \\
\nabla_\mu \nabla_\nu \tau + \frac{1}{2l^2}g_{\mu\nu}(V\circ \tau) &=& 0
\end{eqnarray}
for the dilaton field $\tau(T,r)$ and metric $g$ with scalar curvature $R=R(g)$.  For $\tau(T,r)=\frac{r}{l}$, for example, and a constant $C$ one has the well-known 
solution ~\cite{1,7}
\begin{equation}
 ds^2 = -\left[ -J\left(\frac{r}{l}\right)-C \right]dT^2+\left[
 -J\left(\frac{r}{l}\right)-C \right]^{-1}dr^2
\end{equation}
for $J'(r)=V(r)$.  Actually, $C=-l^2|\nabla \tau|^2 -J \circ \tau$, and $\frac{C}{2l}$ can be interpreted as the energy of the solution.  For spherically symmetric gravity, 
for example, with $V(r)=-\frac{\gamma}{\sqrt{r}}$, $\gamma >0$, one
takes $l=$ the Planck length $l_P$.

For $g$ given by
\begin{equation}
ds^2 = \cos ^2 \frac{u(x,t)}{2} dx^2 - \sin ^2  \frac{u(x,t)}{2} dt^2
\end{equation}
and for $\Delta = \frac{\partial ^2}{\partial x^2}+ \frac{\partial^2}{\partial t^2}$, 
\begin{equation}
 R=\frac{2\Delta u}{\sin u}
\end{equation}
by our sign convention for the scalar curvature, which is opposite the sign
of $R$ in the references ~\cite{3,5}, and the equations in (3) are
\begin{eqnarray}
\tau_{xt}+\frac{1}{2}\tan\left(\frac{u}{2}\right)u_t\tau_x-\frac{1}{2}\cot\left(\frac{u}{2}\right)u_x\tau_t\,&=&0\\
\tau_{xx}+\frac{1}{2}\tan\left(\frac{u}{2}\right)u_x\tau_x+\frac{1}{2}\cot\left(\frac{u}{2}\right)u_t\tau_t+\frac{1}{2l^2}\cos^2\left(\frac{u}{2}\right)(V\circ\tau)\,&=&0\\
\tau_{tt}-\frac{1}{2}\tan\left(\frac{u}{2}\right)u_x\tau_x-\frac{1}{2}\cot\left(\frac{u}{2}\right)u_t\tau_t-\frac{1}{2l^2}\sin^2\left(\frac{u}{2}\right)(V\circ\tau)\,&=&0.
\end{eqnarray}
By equations (2), (6) and by addition of equations (8), (9) on obtains
\begin{eqnarray}
\Delta u &=& -\frac{1}{2l^2}(V'\circ\tau)\sin u \\
\Delta \tau &=& -\frac{1}{2l^2}(V\circ \tau)\cos u,
\end{eqnarray}
equation (10) being a generalized sine-Gordon equation.  Following
M. Cadoni ~\cite{3}, we shall be interested in static field solutions
$u(x,t)=u(x)$, $\tau(x,t)=\tau(x)$ in which case (10), (11) reduce to
the system
\begin{eqnarray}
u''(x) &=& -\frac{1}{2l^2}V'(\tau(x))\sin u(x) \\
\tau ''(x) &=& -\frac{1}{2l^2}V(\tau(x))\cos u(x),
\end{eqnarray}
which has first integrals
\begin{eqnarray}
u'(x) &=& -\frac{A}{l}V(\tau(x)) \\
\tau '(x) &=& \frac{1}{2lA}\sin u(x),
\end{eqnarray}
for any constant $A\neq 0$.  Note that by (14), (15) one easily
deduces that 
\begin{equation}
\frac{d}{dx}\left[ A^2J(\tau(x)))+\sin ^2 \frac{u(x)}{2} \right] =0
\end{equation}
for $J'(x)=V(x)$, which gives the conservation law
\begin{equation}
 A^2J(\tau(x))+\sin ^2 \frac{u(x)}{2}= \mbox{a constant},
\end{equation}
which we express as
\begin{equation} 
\sin ^2 \frac{u(x)}{2} = -A^2 \left[J(\tau(x))+C \right]
 \end{equation}
for a constant C.  Using equations (15), (18) one can also deduce that 
\begin{equation} 
\tau '(x)^2 = -\frac{1}{l^2}\left[J(\tau(x))+C
  \right]\left\{1+A^2\left[J(\tau(x))+C \right] \right\}.
\end{equation}
Equations (18), (19), which we have deduced by a conservation law,
compare with equations (16), (17) in ~\cite{3} where $-K$, $-V$, and $\Phi$
there are our $J$, $V$, and $\tau$, respectively.

\section{The Transformations $\Psi$}
In the static case under consideration we write the metric in (5) as
\begin{equation}
 ds_{sol}^2 =\cos ^2 \frac{u(x)}{2}dx^2-\sin ^2 \frac{u(x)}{2}dt^2,
\end{equation}
where the subscript ``sol'' suggests the word soliton --given equation
(10).  We look for an explicit map $\Psi=(\psi _1 ,\psi_2 )$ and its
inverse $\Theta=(\theta_1,\theta_2)$ such that under the change of
variables $x=\psi_1(T,r)$, $t=\psi_2(T,r)$, the metric $ds_{sol}^2$ in
(20) is transformed to the metric
\begin{equation}
 ds_{bh}^2 = -\left[-J\left(\frac{r}{l}\right)-C
 \right]dT^2+\left[-J\left(\frac{r}{l}\right)-C \right]^{-1}dr^2
\end{equation}
in (4) for the C in equations (18), (19), where the subscript ``bh'' suggests
now some kind of generic extremal black hole.  Conversely, under the
change of variables $T=\theta_1(x,t)$, $r=\theta_2(x,t)$, $ds_{bh}^2
\longrightarrow ds_{sol}^2$.  It turns out that, in contrast to the
more difficult non-static case, $\Psi$ and $\Theta$ can be chosen to
assume the following somewhat simple form:
\begin{align}
\psi_1(T,r) & = \tau^\inv\left(\frac{r}{l}\right),& \psi_2(T,r)  &= 
\frac{T-\theta_0}{A} \\
\theta_1(x,t)& = At+\theta_0,& \theta_2(x,t) & =  l\tau(x) 
\end{align}
for $u(x)$ in (20) that satisfies equation (14) and $\tau(x)$ that
satisfies equation (15); hence $u(x)$ and $\tau(x)$ will solve the
system (12) and (13).  $\tau^\inv$ is the inverse function of $\tau$
and $\theta_0$ is any constant.  The verification that the
transformations in (22), (23) indeed do work relies heavily on the
equations (18), (19), which as we have seen are implied by equations
(14), (15).

As a simple, but important example, choose $V(x)=-2x$ (which provides
for the Jackiw-Teitelboim model), and choose
\begin{equation}
 u(x) = \pm 4 \arctan e^{\frac{x-x_0}{l}}, \,\,\,\,\, \tau(x)=\pm
 \mbox{sech} \left(\frac{x-x_0}{l}\right)
\end{equation}
which solve equations (14), (15) for $A=1$ (and which therefore solve
the sine-Gordon equation $u''(x)=\frac{1}{l^2} \sin u(x)$ (12) and
equation (13)).  Equation (18) holds if and only if $C=0$.  By equation
(22), the transformation of variables
\begin{eqnarray}
x=\psi_1(T,r) &=& x_0+l\,\log\left[ \frac{l+\sqrt{l^2-r^2}}{\pm r}\right] \\
t=\psi_2(T,r) &=& T,
\end{eqnarray}
(where we choose $\theta_0=0$) takes the soliton metric $ds_{sol}^2$
in (20) (for $u(x)$ given in (24)) to the extremal black hole metric
$ds_{bh}^2=-\frac{r^2}{l^2}dT^2+\frac{l^2}{r^2}dr^2$; here
$J(x)=-x^2$.  Similarly by (23) under the transformation of variables
$T=\theta_1(x,t)=t$, $r=\theta_2(x,t)=\pm l\mbox{sech}\left(\frac{x-x_0}{l}\right)$,
$ds_{bh}^2 \longrightarrow ds_{sol}^2$.  This transformation and its
inverse $\Psi$ in (25), (26) can be seen as implementing the Cadoni
correspondence for the present example.  Note a minor typing error in
equation (31) of ~\cite{3}: there one should have $\Phi^\inv =\pm
\cosh(\lambda(x-x_0))$ (instead of $\Phi^\inv =\cosh(\lambda(x-x_0))$,
as the minus sign is needed for the minus in (24)).

As pointed out in ~\cite{3}, a complete correspondence between 2D space-time
structures of the dilaton gravity theory and solutions in generalized
sine-Gordon field theory requires a consideration of the metric (5)
and of the metric
\begin{equation}
ds^2 =\cosh ^2 \frac{u(x,t)}{2}dx^2-\sinh ^2 \frac{u(x,t)}{2}dt^2
\end{equation}
as well, for the sinh-Gordon model.  Here for $\square =
\frac{\partial ^2}{\partial x^2} - \frac{\partial ^2}{\partial t^2}$,
$R=\frac{2\square u}{\sinh u}$ and the equations (2), (3) are now
\begin{eqnarray}
\square u + \frac{1}{2l^2}(V' \circ \tau) \sinh u &=& 0 \\
\tau_{xt}-\frac{1}{2}\tanh\left(\frac{u}{2}\right)u_t\tau_x-\frac{1}{2}\coth\left(\frac{u}{2}\right)u_x\tau_t\,&=&0\\
\tau_{xx}-\frac{1}{2}\tanh\left(\frac{u}{2}\right)u_x\tau_x-\frac{1}{2}\coth\left(\frac{u}{2}\right)u_t\tau_t+\frac{1}{2l^2}\cosh^2\left(\frac{u}{2}\right)(V\circ\tau)\,&=&0\\
\tau_{tt}-\frac{1}{2}\tanh\left(\frac{u}{2}\right)u_x\tau_x-\frac{1}{2}\coth\left(\frac{u}{2}\right)u_t\tau_t-\frac{1}{2l^2}\sinh^2\left(\frac{u}{2}\right)(V\circ\tau)\,&=&0,
\end{eqnarray}
from which one obtains in the static case the system
\begin{eqnarray}
u''(x) &=& -\frac{1}{2l^2}V'(\tau(x))\sinh u(x) \\
\tau ''(x) &=& -\frac{1}{2l^2}V(\tau(x))\cosh u(x)
\end{eqnarray}
with first integrals
\begin{eqnarray}
u'(x) &=& -\frac{A}{l}V(\tau(x)) \\
\tau '(x) &=& \frac{1}{2Al}\sinh u(x),
\end{eqnarray}
that compare with equations (14), (15).  Equations (18), (19) are
replaced by
\begin{eqnarray}
\sinh ^2 \frac{u(x)}{2} &=& -A^2 \left[J(\tau(x))+C \right] \\
\tau '(x)^2 &=& -\frac{1}{l^2}\left[J(\tau(x))+C \right]\left\{1-A^2\left[J(\tau(x))+C \right] \right\}
\end{eqnarray}
for a suitable constant $C$.  For this $C$, and for $u(x)$, $\tau(x)$,
that solve (34), (35) (hence $u(x)$, $\tau(x)$ also solve equations (32),
(33) ) one can check that for the metric $ds^2$ in (27), $ds^2
\longrightarrow ds_{bh}^2$ in (21), under the change of variables
$(x,t) \longrightarrow \Psi(T,r)=(\psi_1(T,r), \psi_2(T,r))$, where
\begin{align}
\psi_1(T,r) & =  \tau^\inv\left(\frac{r}{l}\right),& \psi_2(T,r) & = 
\frac{T-\theta_0}{A} \\
\theta_1(x,t)& =  At+\theta_0,& \theta_2(x,t) & =  l\tau(x); 
\end{align}
here $\Theta=(\theta_1, \theta_2)=\Psi^\inv$.  Thus $\Psi$, $\Theta$
here have the same form as the $\Psi$, $\Theta$ in (22), (23).

\section{The Sinh-$\Phi$ model and other examples}
Another model is defined by the potential $V(x)=-\sinh(2x)$.
Equations (14), (15) are solved by
\begin{eqnarray}
u(x) &=& \pi + 2 \arctan \left[ \sqrt{2} \sinh\left( \frac{x-x_0}{l}\right)\right] \\
\tau(x) &=& \mbox{arctanh} \left[\frac{1}{\sqrt{2}}\mbox{sech}\left(\frac{x-x_0}{l}\right)  \right]
\end{eqnarray}
for $A=1$, and equation (18) holds for $C=\frac{1}{2}$.
$J(x)=-\frac{1}{2}\cosh(2x)$.  The extremal black hole solution
$ds_{bh}^2$ corresponding to the solutions (40), (41), in the Cadoni
correspondence, is given by equation (21):
\begin{eqnarray}
ds_{bh}^2 &=&
-\left[\frac{1}{2}\cosh\left(\frac{2r}{l}\right)-\frac{1}{2}
  \right]dT^2+\left[\frac{1}{2}\cosh\left(\frac{2r}{l}\right)-\frac{1}{2} \right]^{-1}dr^2 \nonumber\\
 &=& -\sinh ^2 \left( \frac{r}{l}\right)dT^2+ \sinh ^{-2} \left( \frac{r}{l}\right)dr^2,
\end{eqnarray}
as in equation (40) of ~\cite{3}.  A change of variables $x=\psi_1(T,r)$,
$t=\psi_2(T,r)$ that takes $ds_{sol}^2$ in (20) (for $u(x)$ in (40))
directly to $ds_{bh}^2$ in (42) is given by equation (22), where we
note that $\tau^\inv(x) =
x_0+l\mbox{arcsech}\left(\sqrt{2}\tanh(x)\right)$ for $\tau(x)$ in
(41):
\begin{eqnarray}
\psi_1(T,r) &=& x_0 + l\,\log \left[ \frac{1+\sqrt{1-2\tanh ^2 \frac{r}{l}}}{\sqrt{2}\tanh\frac{r}{l}} \right] \\
\psi_2(T,r) &=& T,
\end{eqnarray}
where again we take $\theta_0=0$.

Going back to the Jackiw-Teitelboim model with $V(x)=-2x$, one can
obtain another solution $u(x)$, $\tau(x)$ of the field equations (12),
(13) - one of independent interest that involves the Jacobi elliptic
funcitons sn, cn, and dn ~\cite{8}.  For this, given $A\neq 0$ (as in
equations (14), (15)) and a constant $E>0$, define
\begin{eqnarray}
B &=& \frac{A^2E}{4l^2}\nonumber \\
K &=& 1+\frac{2}{A^2E}\nonumber\\
\alpha &=&  \sqrt{2K^2+2K\sqrt{K^2-1}-1}\\
g(x) &=& \left(\sqrt{B}\sqrt{\sqrt{K^2-1}-K}\right)x\nonumber \\
f(x) &=& \frac{\mbox{sn}(g(x),
  \alpha)}{\sqrt{\sqrt{K^2-1}-K}}\nonumber .
\end{eqnarray}
Then one can show that the pair
\begin{eqnarray}
u(x) &=& 4 \arctan f(x)\\
\tau(x) &=& \frac{\sqrt{E} \mbox{cn}(g(x),\alpha)\mbox{dn}(g(x),\alpha)}{1+\frac{\mbox{sn}^2(g(x),\alpha)}{\sqrt{K^2-1}-K}}
\end{eqnarray}
solves equations (14), (15).  Since $\mbox{sn}(0,\alpha)=0$ and
$\mbox{cn}(0,\alpha)=\mbox{dn}(0,\alpha)=1$, we see that $u(0)=0$ and
$\tau(0)=\sqrt{E}$.  Also, $J(x)=-x^2$, which means that in equation
(18) we can conclude that $C=\tau(0)^2=E$, and that the solution in
(21) corresponding to (46), (47) is given by
\begin{equation}
 ds_{bh}^2 = -\left[\frac{r^2}{l^2}-E \right]dT^2+
 \left[\frac{r^2}{l^2}-E \right]^{-1}dr^2,
\end{equation} 
which is a black hole with positive mass $E$.  In this case it is
easier to compute the inverse tranformation $\Theta =\Psi ^\inv$ in
(23) which is simply $l\tau$ for $\tau$ in (47).

The string-inspired gravity model, with $V(x)=-\gamma$, $\gamma>0$,
gives rise to the (non-soliton) example
\begin{equation}
u(x)=\frac{A\gamma}{l}x+b, \,\,\,\,\,
\tau(x)=-\frac{1}{2A^2\gamma}\cos\left(\frac{A\gamma}{l}x+b\right)+c
\end{equation}
that solves (14, (15).  $J(x)=-\gamma x$ and the choice of
$x=-\frac{bl}{A\gamma}$ in (18) gives $C=-\frac{1}{2A^2}+\gamma c$, by
which one obtains the solution (see (21))
\begin{equation}
 ds_{bh}^2 = -\left[\frac{\gamma}{l}r +\frac{1}{2A^2}-\gamma
 c\right]dT^2+\left[\frac{\gamma}{l}r +\frac{1}{2A^2}-\gamma
 c\right]^{-1}dr^2.
 \end{equation}
$\Psi(T,r)$ in (22) assumes the form
\begin{eqnarray}
\psi_1(T,r) &=& \frac{l}{A\gamma}\left\{-b+\arccos\left[ 2A^2\gamma \left(c-\frac{r}{l}\right)\right]\right\} \\
\psi_2(T,r) &=& \frac{T-\theta_0}{A}. 
\end{eqnarray}

\section{Remarks on non-static solutions}
Non-static solutions of the field equations (7), (8), (9) (for
string-inspired, spherically symmetric, and Jackiw-Teitelboim (JT)
gravity) are given in ~\cite{2}, that complement 2-soliton solutions found
in ~\cite{5} for JT gravity.  For example, given $m$ and $v>0$ (which we
regard as mass and velocity parameters) set $a^2=1+v^2$ and define
\begin{eqnarray}
u(x,t) &=& 4 \arctan \left[\frac{v \sinh(amx)}{a\cos(vmt)} \right]\\
\tau(x,t) &=& \frac{4v^2am\left[\sin(vmt)\right]\sinh(amx)}{a^2\cos
  ^2(vmt) + v^2 \sinh ^2(amx)}.
\end{eqnarray}
This pair solves the system (7)-(9) for $V(x)=-x$,
$l=\frac{1}{\sqrt{2}m}$, as shown in ~\cite{2}, where a transformation
$\Psi$ is also constructed that takes the metric in (5) to that in
(4), for an appropriate value of $C$.  Another application for such
transformations $\Psi$ is toward the construction of exact solutions
of field equations defined by the Laplacian $\square_{sol}^{+}$ of the metric
(5).  In the static case at hand, for example, one can prove the following
commutivity: let $\square_{sol}^{-}$ denote the Laplacian of the metric in
equation (27), let $\square_{bh}$ denote the Laplacian of the metric in
equation (4), and let $D_\tau^{\pm}$ denote the domains
\begin{equation}
D_\tau^\pm = \{ (x,t) \in \mathbb{R}^2 | \pm \tau '(x)>0 \}.
\end{equation}
Then for a function $f(T,r)$ and the transformation $\Theta$ in (23) (which
we have observed is the same as that in (39)), one has
\begin{eqnarray}
\square_{sol}^+(f\circ\Theta) &=& \left( \square_{bh} f\right)\circ\Theta
\quad \mbox{on }\,\,D_\tau^+\nonumber\\
\square_{sol}^-(f\circ\Theta) &=& \left( \square_{bh} f\right)\circ\Theta
\quad \mbox{on }\,\,D_\tau^-\,\,.
\end{eqnarray}
This is based on the somewhat remarkable property that these
$\Theta=\Psi^{-1}$, in principle, commute with $\square_{sol}$ and
$\square_{bh}$, which means that they
are also transformations of solution spaces.  Details of this
commutation property are found in ~\cite{11}, in the JT case, where some
solutions of the equation $\square_{sol} \phi = \mu \phi$ are also
presented.


\begin{thebibliography}{99}

\bibitem{1} T. Banks and M. O'Laughlin, {\it Nucl. Phys.} {\bf B362}, 649 (1991).

\bibitem{2} S. Beheshti, Ph.D. Thesis, University of Massachusetts,
  Amherst, MA (2006).

\bibitem{3} M. Cadoni, {\it Phys. Rev. D58} {\bf 104001}, (1998).

\bibitem{4} J. Gegenberg and G. Kunstatter, {\it Phys. Lett.} {\bf B413}, 274 (1997).

\bibitem{5} J. Gegenberg and G. Kunstatter, {\it Phys. Rev. D58} {\bf 124010}, 274 (1998).

\bibitem{6} J. Gegenberg and G. Kunstatter, from {\it Solitons: Properties, Dynamics, Interactions, Applications, M. Mackenzie, R. Paranjape, and W. Zakrzewski, Editors.} {\bf Chapter 14}, 99 (2000).

\bibitem{7} D. Louis-Martinez, J. Gegenberg, and G. Kunstatter, {\it
  Phys. Lett.} {\bf B321}, 193 (1994).

\bibitem{8} H. Hancock, {\it Lectures on the Theory of Elliptic
  Functions}, Dover Pub. (1958).

\bibitem{9} F. Williams, from {\it Quantum Field Theory Under the
  Influence of External Conditions, K. Milton, Editor}, 370, Rinton Press (2004).

\bibitem{10} F. Williams, from {\it Proceedings of the Fourth International
Conference on Mathematical Physics, Rio de Janeiro, Brazil} (2004),
  Proceedings of Science, http://pos.sissa.it.

\bibitem{11} F. Williams, from {\it Trends in Soliton Research,
  L. Chen, Editor}, {\bf Chapter 1}, 1 Nova Science Pub. (2006).

\end{thebibliography}
\end{document}